\documentclass[12pt]{article}
\usepackage{amsfonts,amssymb,amsmath}
\usepackage{graphics}
\usepackage{lscape}
\usepackage{makeidx}
\usepackage{theorem}
\usepackage[dvips]{graphicx}

\begin{document}

\newpage
\pagestyle{empty}


\rightline{DSF/14/2008}


\begin{center}

    {\Large \textbf{\textsf{Arithmetical Chaos and Quantum Cosmology}}}

    \vspace{10mm}

    {\large L. A. Forte}

    \vspace{10mm}

    \emph{Dipartimento di Scienze Fisiche, Universit{\`a} di Napoli
    ``Federico II''}

    \emph{and I.N.F.N., Sezione di Napoli,}

    \emph{Complesso Universitario di Monte S. Angelo,}

    \emph{Via Cinthia, I-80126 Naples, Italy}
    \end{center}

\begin{abstract}
In this note, we present the formalism to start a quantum analysis
for the recent billiard representation introduced by Damour,
Henneaux and Nicolai in the study of the cosmological singularity.
In particular we use the theory of Maass automorphic forms and
recent mathematical results about arithmetical dynamical systems.
The predictions of the billiard model give precise automorphic
properties for the wave function (Maass-Hecke eigenform), the
asymptotic number of quantum states (Selberg asymptotics for
PSL$(2,\mathbb{Z})$), the distribution for the level spacing
statistics (the Poissonian one) and the absence of scarred states.
The most interesting implication of this model is perhaps that the
discrete spectrum is fully embedded in the continuous one.
\end{abstract}

      \vfill

    \vfill
    \vspace*{3mm}

    \hrule

\vspace*{2mm} \emph{E-mail:} \texttt{forte@na.infn.it}
    \vspace*{3mm}

    \newpage
    \pagestyle{plain}
    \setcounter{page}{1}
    \baselineskip=16pt
    \parindent=0pt

\section{Introduction}

The beautiful singularity theorems of S. Hawking and R. Penrose
\cite{HawkingPenrose} establish the presence of a singularity in
the generic solution to Einstein field equations under very
reasonable assumptions of the kind of matter. These theorems use
tools from differential topology (Morse theory for Lorentzian
manifolds) but do not say anything about the behavior of the
singularity, being existence results. Thus the works by V.A.
Belinskii, I.M. Khalatnikov, E.M. Lifshitz (BKL in the following)
\cite{BKL} regarding the general behavior of a homogeneous
cosmological singularity are very important in understanding the
(space-like) singularity in general relativity. Their approach is
based on the assumption that close to the cosmological singularity
the spatial gradients can be neglected with respect to the
temporal ones in the field equations, this being equivalent to a
spatial decoupling of the points on the singular hyper-surface
(BKL limit). Thus each point (i.e. the space part of the metric)
evolves according to an infinite succession of Kasner eras,
leading to an approximate discretized dynamics which turns out to
be isomorphic to the Gauss map (the continued fraction expansion
of a real number), hence chaotic. The analysis of BKL has been
recently improved and generalized by T. Damour, M. Henneaux and H.
Nicolai (DHN in the following) \cite{DHN} to the case of a general
inhomogeneous cosmological singularity and supergravity/string
theories (extra dimensions, exotic matter etc). Their result is
that, in the BKL limit, the dynamics of pure gravity in 4
dimensions can be reformulated as a billiard motion in a suitable
domain embedded in a flat 3-dimensional pseudo-Riemannian manifold
(we comment on the chaoticity of this billiard too). This billiard
table has a basis on the hyperbolic plane corresponding to the
fundamental domain of the extended modular group
PGL$(2,\mathbb{Z})$, which is a so-called arithmetic group. The
arithmetic nature of the asymptotic billiard allows to derive a
precise quantum analysis and make some predictions for this model
using powerful and recent theorems about the quantum behavior of
arithmetical dynamical systems. This quantum analysis is the main
point of this note. In particular, we can give the precise form of
the wave function, count the asymptotic number of quantum states
and comment on the so-called scarred states. Finally, the role of
the Selberg trace formula for PSL$(2,\mathbb{Z})$ as a convergent
semi-classical quantization rule for gravity in this regime is
briefly pointed out.

Two appendices, on the Kac-Moody algebra HA$_{1}^{(1)}$ and the
theory of Maass waveforms, are added at the end and should be read
in parallel with the relevant sections.

This paper is based on \cite{Forte}, which also contains more
details, references and background material. For the BKL limit,
the original references and many other related questions we
suggest the recent reviews \cite{Montani} and \cite{Henneaux}; the
latter deals especially with the cosmological billiards in
relation to other theories (supergravity etc) too.

We want to stress that, usually, with the term chaotic cosmology
one refers to the fact the asymptotic (i.e. in the limit towards
the singularity) classical evolution exhibits some chaotic
features, but one has to consider what are the implications of
that in the quantum behavior of the system. As we will see the
corresponding quantum system is not ``chaotic'', since the quantum
manifestations of classical chaos for arithmetical systems are
closer to a classically integrable system, and this is due to the
arithmetic nature of the billiard. In fact, the existence of Hecke
operators, which form at the quantum level an infinite family of
commuting self-adjoint operators which mimic an integrable system,
allows for an anomalous spectral statistics (a Poissonian law for
the high energy spectrum of the quantum Hamiltonian instead of a
GOE/GUE ensemble) and allows to derive the quantum unique
ergodicity theorem.

\section{The BKL oscillatory behavior and the billiard representation}

Let us briefly review the BKL result, restricting to the case of a
homogeneous Bianchi IX universe, or Mixmaster as renamed by C.
Misner for the chaotic features it exhibits. Moreover, we consider
only the case of pure gravity in 4 dimensions. Then, the spacetime
metric can be written in a synchronous reference system as
\begin{equation}\label{bianchimetric}
ds^{2} = -dt^{2} + dl^{2} = -dt^{2} + h_{ij}dx^{i}dx^{j}
\end{equation}
with $h_{ij}$ diagonal
\begin{equation}
h_{ij} = a^{2}(t)l_{i}l_{j} + b^{2}(t)m_{i}m_{j} +
c^{2}(t)n_{i}n_{j}
\end{equation}
and where the frame vectors $l,m,n$ depend on the spatial
coordinates $x^{i}$. The behavior of the three scale factors
$a,b,c$ has been studied by BKL in the limit $t \rightarrow
0$\footnote{Regarding the validity of the BKL regime, we have to
say that it should be settled down in the pre-inflationary
Universe, thus unaccessible to present observation. Thus it
concerns the very early stages of the Universe evolution, matching
the Planckian era with the inflationary behavior. Of course, one
should always consider the problem of the falsification of a
theory of quantum gravity, since it deals with Planck scale
physics.}. Their evolution is represented in the picture.
\begin{figure}[htbp]
\centering
  \includegraphics[width=5cm,keepaspectratio]{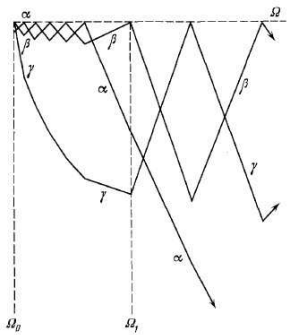}\\
  \caption{Evolution of the logarithms of the scale factors $(\alpha=\ln a, \beta=\ln b, \gamma = \ln c)$
  in terms of a logarithmic time $\Omega= - \ln t$ for
  a Bianchi IX universe (Mixmaster).}
\end{figure}
The asymptotic dynamics is such that there exist an infinite
number of Kasner eras, each one formed by a certain number of
Kasner epochs. Precisely, during each \emph{Kasner era}, two of
the scale factors oscillate and the third one decreases
monotonically. On passing from one era to the next one, the
monotonic decrease is transferred to another of the three scale
factors. The time interval in which each scale factor is
represented as a straight segment is called a \emph{Kasner epoch},
i.e. it is a regime of the solution to the field equation where
one can approximate the exact solution with a Kasner solution.
Given an irrational number $u > 1$, the number of Kasner epochs
contained in a Kasner era corresponds to the partial denominators
in the continued fraction expansion for $u$.

In other words, the result is that the asymptotic evolution can be
described as infinite succession of Kasner epochs with a certain
law of replacement of the Kasner exponents when passing over from
one epoch to the next one. This law corresponds to the Gauss map
\begin{equation}
T(x) = \frac{1}{x} -
\left[\frac{1}{x}\right]=\left\{\frac{1}{x}\right\}
\end{equation}
whose metric entropy can be computed
\begin{equation}
h(T_{Gauss}) = \frac{\pi^{2}}{6 \ln 2}
\end{equation}
Note that the Gauss map accounts only for the transitions between
successive Kasner eras (i.e. in the picture between the times $\Omega_{0}$ and $\Omega_{1}$ and so on).
One can do better and consider also the
oscillations in two of the scale factors inside a single Kasner
era and include them in a more complete map which describes the
discrete evolution (still approximated). This can be realized
through the Farey map, whose topological entropy is given by
\begin{equation}
\tau(T_{Farey}) = 2\,\ln 2
\end{equation}
This map accounts for oscillations (one pair of axes oscillates
while the third one decreases monotonically) and bounces (when the
role of the three axes are interchanged and a different axis
decreases monotonically), according to a chaotic Farey tale
\cite{Levin}. This paper uses fractal techniques (which are
observer independent) to claim that the Mixmaster universe is
indeed chaotic.

Finally, note that in the BKL approach (although fascinating) it
is not easy to write down explicitly a quantization procedure.

BKL also considered the case of a inhomogeneous universe and
showed that there exist a generalized Kasner solution near the
singularity
\begin{eqnarray}
  dl^{2} &=& h_{ij}\,dx^{i}dx^{j} \\
  h_{ij} &=& a^{2}l_{i}l_{j} + b^{2}m_{i}m_{j} + c^{2}n_{i}n_{j}
\end{eqnarray}
with
\begin{equation}
a \sim t^{p_{l}},\quad b \sim t^{p_{m}},\quad c \sim t^{p_{n}}
\end{equation}
and
\begin{equation}
p_{l}(x^{i}) + p_{m}(x^{i}) + p_{n}(x^{i}) = p_{l}^{2}(x^{i}) +
p_{m}^{2}(x^{i}) + p_{n}^{2}(x^{i}) = 1
\end{equation}
but now, differently from the homogeneous case, the Kasner
exponents are functions of the spatial coordinates. One expresses
this fact by saying that inhomogeneous cosmology can be modelled
on a Bianchi IX universe, since there exists an analog infinite
succession of Kasner epochs/eras with local behavior, i.e. each
point on the singular hypersurface evolves with the same laws
(Bianchi IX) but with different parameters. In this sense, one
says that the behavior of a generic singularity is \emph{local}
and \emph{oscillatory}.

The positivity of the metric entropy for the Gauss map implies its
ergodicity. This result was derived for the first time by Artin
(much before the introduction of entropy in dynamical systems)
relating the properties of the Gauss map to the geodesic flow on
PSL$(2,\mathbb{Z})\backslash\mathbb{H}$, which was known to be
ergodic (see \cite{Sinai}). Thus, one can ask if there exists a
precise physical system associated to this geodesic flow in the
same way as the asymptotic evolution of a Bianchi IX universe can
be approximately described by the Gauss map. The answer to this
question is positive and the physical system is pure gravity in 4
dimensions near the singularity \emph{and} without any symmetry
assumption for the metric. This leads us to the billiard
representation introduced by DHN.

The result by DHN \cite{DHN} is that the BKL oscillatory behavior
of pure gravity in 4 dimensions can be reformulated as a billiard
motion in an auxiliary spacetime\footnote{One can speculate
whether this representation is photographic or holographic (since
we go from a 4-dimensional spacetime to a 3-dimensional auxiliary
Minkowski space), according to the words of T. Damour at the last
11th Marcel Grossmann meeting.}. More precisely, the dynamics of
the generic solution to Einstein equations close to the
cosmological singularity (and in the BKL limit, i.e. assuming
BKL's conjecture) is equivalent to a \emph{null} geodesic motion
(i.e. on a light ray) inside a billiard table given by a Coxeter
polytope in a 3-dimensional Minkowski space\footnote{Generic means
without any symmetry assumption for the metric.If accidental
symmetries are present in the metric, the analysis is \emph{not}
valid. For example, it is known that the Schwarzschild solution
\begin{equation}
ds^{2} = -\left(1-\frac{2m}{r} \right)dt^{2} +
\left(1-\frac{2m}{r} \right)^{-1}dr^{2} + r^{2}d\Omega^{2}
\end{equation}
has a spacelike singularity at $r=0$; moreover, inside the horizon
$(r<2m)$ the $r$ coordinate is time-like (there is a minus sign in
front of $dr^{2}$). If we take the limit $r \rightarrow 0$, we
obtain
\begin{equation}
\lim_{r \rightarrow 0} ds^{2} = \frac{2m}{r}\,dt^{2}  -
\frac{r}{2m}\,dr^{2} + r^{2}\,d\Omega^{2}
\end{equation}
which is a Kasner metric
\begin{equation}
-d\tau^{2} + \tau^{-2/3}d\sigma^{2} + \tau^{4/3}(\sin(\theta)
d\overline{\phi})^{2}
\end{equation}
once we put $\tau = \frac{2r^{3/2}}{3\sqrt{2m}}$,
$\sigma=(4m/3)^{1/3}t$, $\overline{\theta} = (9m/2)^{1/3}\theta$
and $\overline{\phi} = (9m/2)^{1/3}\phi$, i.e. the Schwarzschild
solution corresponds (in the neighbourhood of the singularity) to
a \emph{single} Kasner epoch, not to a never-ending succession of
Kasner eras.

Finally, the analysis does not apply to time-like (like the one in
the Reissner-Nordstr\"{o}m solution, the charged black hole) or
null singularities where a causal decoupling of spatial points
does not occur. The question about the general behavior of
not-space-like singularities is still open.}. The remarkable fact
is that this polytope corresponds to the Weyl chamber of the
hyperbolic Kac-Moody algebra HA$_{1}^{(1)}$, the canonical
hyperbolic extension of A$_{1}$ ($\mathfrak{su}(2)$). The walls
are the hyperplanes pointwise fixed by Weyl reflections in the
three simple roots which define the algebra and the reflections at
the walls are elastic. In this language, each Kasner epoch is
represented by a null geodesic segment between two successive
reflections. In particular, given a Kasner epoch and the wall
where this epoch crashes/ends, the following one is obtained by
Weyl reflection with respect to the simple root orthogonal to that
face of the Weyl chamber. In other words, Weyl reflections with
respect to simple roots send null geodesic segments into null
geodesic segments, i.e. transform Kasner solutions into Kasner
solutions (with different values of Kasner exponents). Because of
this, we have a hint that the Weyl group of the algebra acts on
the space of solutions of Einstein equations, transforming Kasner
solutions into Kasner solutions which can then be associated to
the simple roots (and more generally to the real roots of the
algebra). Of course, this would be a technique to generate
solutions (also approximate) to Einstein equations starting from
very simple solutions. This was already done by R. Geroch, who
found that gravity reduced to 1+1 dimensions has a hidden
A$_{1}^{(1)}$ symmetry (affine $\mathfrak{su}(2)$), which is a
sub-algebra of HA$_{1}^{(1)}$. Thus in this case, the affine
algebra (or better its corresponding infinite-dimensional Lie
group, known as the Geroch group) can be considered a symmetry of
the theory. One can go further and check if the whole hyperbolic
algebra (not only its Weyl group, on which we focus in this work)
is a symmetry of gravity and in which regime, see \cite{Henneaux}.

In this formalism only null geodesics are physical and correspond
to Kasner solutions. Space-like and time-like geodesics seem to
play no role. Note that the walls of the Weyl chamber are
time-like, that is their orthogonal vectors are space-like, since
for the simple roots one has $(\alpha_{i},\alpha_{i})=2 > 0$.
Because of this, every reflection preserves the null character of
the velocity vector.

The (flat) metric in this Minkowski space is given by the Cartan
matrix of the algebra
\begin{equation}
A=\left(%
\begin{array}{ccc}
  2 & -1 & 0 \\
  -1 & 2 & -2 \\
  0 & -2 & 2 \\
\end{array}%
\right)
\end{equation}
whose signature is $(-++)$. For this reason, one has three kinds
of geodesics, but only \emph{null} geodesics appear.

Note that in the DHN approach, the 3-dimensional billiard is
peculiar, in fact the walls move \emph{together} with the
trajectories. During this motion the geometry of the billiard
remains the same, i.e. the structure of the algebra is preserved
and the Weyl group is uniquely determined. This means that the
walls of the (reduced) billiard on the hyperbolic plane are
motionless and the 2-dimensional analysis is much easier than
studying the 3-dimensional billiard with (say) co-moving walls. As
we will say many times in this work, the Weyl group of the algebra
is an arithmetic group (being isomorphic to PGL$(2,\mathbb{Z})$),
thus we are coping with a physical system represented by an
arithmetical dynamical system.

The very essence of arithmetical chaos (see \cite{BGS}) must be
researched in its quantum aspects, since from the classical point
of view no difference exists between generic chaotic dynamical
systems and arithmetical dynamical systems, except for the
degeneracy of lengths of periodic orbits. In fact, for generic
systems, one does not expect a degeneracy of this kind but the
ones which come from the symmetry of the model. For example, for
time-invariant systems, this multiplicity, on average, is 2, which
corresponds to the two possible ways to run a geodesic.
Arithmetical systems are very exceptional, since the degeneracy of
lengths of periodic orbits grows exponentially
\begin{equation}
\frac{1}{C}\,\frac{e^{L/2}}{L/2}
\end{equation}
where $L$ is the length of such a geodesic and $C$ is a constant
depending of the particular group ($C=1$ for PSL$(2,\mathbb{Z})$).
To this, one should add the Horowitz-Randol theorem which states
that these degeneracies are unbounded. The large degeneracy of
lengths of periodic orbits seems to have no importance in
classical mechanics. These systems are ergodic as any other model
on negatively curved surfaces, but the quantum aspects are
anomalous. Thus, it is very interesting to start a quantum
analysis of the cosmological billiard using its arithmetic nature.
Before doing that, in the next section we comment on the
(classical) chaoticity of the cosmological billiard.

\section{General Considerations on the Cosmological Billiard}

Let us stress again that the billiard motion occurs in a
pseudo-Riemannian 3-dimensional space\footnote{We prefer the term
pseudo-Riemannian instead of Lorentzian because for people who
study billiards as dynamical systems, the word Lorentzian has a
completely different meaning.}, \emph{not} in a Euclidean space.
This space is $\mathfrak{h}_{\mathbb{R}}^{\ast}$, in the notion of
Kac \cite{Kac} (see also Appendix \ref{kacmoody}). Thus the
billiard motion is an interrupted null flow on a pseudo-Riemannian
manifold. This situation is different from the the typical
billiards which are embedded in Riemannian manifolds. In fact, in
pseudo-Riemannian manifolds, one has three kinds of geodesics, and
consequently three kinds of \emph{geodesic} flows (indeed one
first needs to define in a rigorous way flows for manifolds with
an indefinite metric). The second step is the definition of the
\emph{billiard} flow\footnote{The problem is the definition of the
billiard law, i.e. how to use the rule "angle of incidence = angle
of reflection" in a pseudo-Riemannian setting. When one speaks of
billiards, one must be careful, because together with the billiard
table, one has also to specify the billiard law. In Euclidean
spaces, one usually considers the law of geometrical optics, but
in principle one can study different billiard laws (in the same
table), as it happens for \emph{outer billiards} which have
nowadays surprising applications.}, which is even more involved.
In fact, because of the three kinds of directions (null-like,
time-like, space-like), each reflection on a wall depends both on
the character of the incoming trajectory and on the character of
the wall.

\begin{figure}[htbp]
\centering
  \includegraphics[width = 6cm,keepaspectratio]{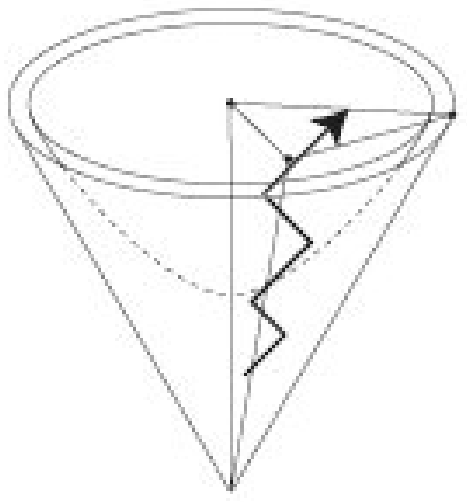}%
   \includegraphics[width = 6cm,keepaspectratio]{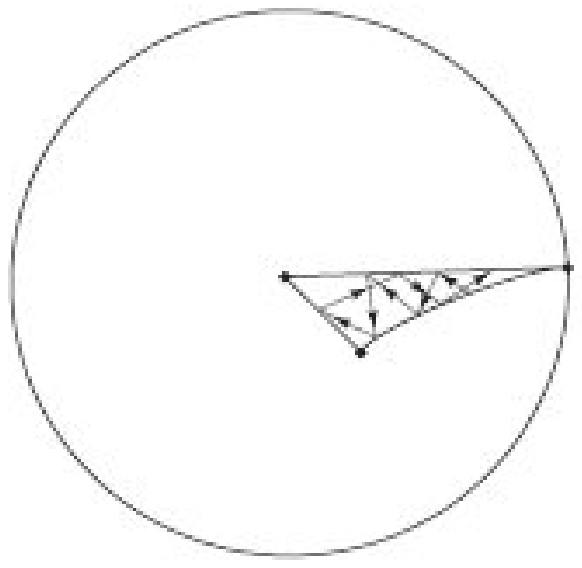}
  \caption{Schematic representation for the cosmological billiard in the Weyl chamber of HA$_{1}^{(1)}$ and projection
  on the Poincar\'{e} disc.}
\end{figure}

To clarify this important point, let us consider the usual
billiard motion in a convex polyhedron $Q$ in a Euclidean space,
that is a Riemannian manifold with the usual (diagonal) flat
metric (see \cite{Sinai}). The boundary of this polyhedron is
given by the union of its faces $\Gamma_{i}$. Each elastic
reflection in the wall $\Gamma_{i}$ is given by the geometric
reflection $\sigma_{i}$
\begin{equation}
\sigma_{i}(v) = v - 2(n_{i},v)n_{i}
\end{equation}
where $v$ is the velocity vector of the incoming trajectory,
$n_{i}$ the unit vector orthogonal to each face and $(,)$ is the
standard Euclidean scalar product (compare with the expression of
the Weyl reflections in Appendix \ref{kacmoody}). To understand
the properties of the billiard flow inside $Q$, one has to
consider the group $G_{Q}$ generated by the reflections
$\{\sigma_{i}\}$; this is a subgroup of all isometries of the
Riemannian manifold. The ergodicity of the billiard depends on
$G_{Q}$, precisely: if $G_{Q}$ is a \emph{finite} group, then the
billiard flow inside $Q$ is \emph{not} ergodic (thus it is not
chaotic). In 2 dimensions, the finiteness of the group is
equivalent to the commensurability of all angles on the polygon
$Q$.

The situation for generic polyhedra is still open. Indeed, one
knows that the metric entropy of a billiard inside an arbitrary,
\emph{not} necessarily convex, polyhedron is zero. All these
statements are valid for Riemannian manifolds.

In our case, the analogy is the following. The polyhedron $Q$ is
replaced by the Weyl chamber of HA$_{1}^{(1)}$, the Riemannian
manifold is replaced by the pseudo-Riemannian space
$\mathfrak{h}_{\mathbb{R}}^{\ast}$ and the billiard motion occurs
on null segments. The faces of the billiard are timelike walls.
Finally, the reflection group $G_{Q}$ is replaced by the Weyl
group of the algebra $W$, which is an infinite Coxeter group. At
the moment of writing this work (and modulo over-present
ignorance), there are no mathematical results which imply the
ergodicity (or the integrability) of our billiard motion, because
the difficulty is that the motion occurs in pseudo-Riemannian
manifolds, thus we can not invoke the previous theorems for
billiard flows in Riemannian manifolds. In particular, we know
that $W$ is isomorphic to PGL$(2,\mathbb{Z})$, and that the free
motion on the hyperbolic plane inside a region with finite
(hyperbolic) area is a chaotic flow, being an Anosov flow
(\cite{Sinai}). Thus we can state that there exists a chaotic
motion on the ``basis'' of the billiard, but this does \emph{not}
imply the chaoticity of the null billiard flow in the full
3-dimensional Weyl chamber. In fact, the full motion could be less
chaotic or even integrable. For example, the Bunimovich stadium is
a well-known example of a 2-dimensional Euclidean chaotic
billiard, but if we consider a 3-dimensional Euclidean billiard
raising the stadium as a basis in the $z-$direction, then the
billiard is integrable, because of a translation symmetry in the
$z-$direction. Thus, a billiard with a chaotic basis (i.e. which
has the property of being ergodic or mixing or hyperbolic etc) is
not necessarily chaotic. Our case is even more complicated,
because the full 3-dimensional billiard is a Coxeter polytope in a
pseudo-Riemannian space, and the projected billiard corresponds to
the fundamental domain of PGL$(2,\mathbb{Z})$ on the hyperbolic
plane (which is a Riemannian manifold) \footnote{With the
projection of the 3-dimensional billiard on the hyperbolic plane,
we mean that we consider the geodesic segment on the hyperbolic
plane corresponding to the intersection in 3 dimensions of a null
segment with a plane through the origin.}.

To conclude this section, one can not state that the Kac-Moody
billiard is chaotic only from the property that the billiard table
is contained inside the light-cone, but more powerful (and at the
moment missing) theorems are needed. The study of billiard flows
on pseudo-Riemannian manifolds is a completely new field and
should be an interesting topic of future research. It seems that
the only mathematical papers on this subject are very recent and
due to B. Khesin and S. Tabachnikov \cite{KhesinTaba}, where they
define a billiard law which is area-preserving and does not change
the type of a geodesic.

Let us also observe that the metric entropy of the geodesic flow
inside the fundamental domain of PGL$(2,\mathbb{Z})$ can be
explicitly calculated\footnote{In fact, for the geodesic flow
$\{S^{t}\}$ on a surface of negative constant curvature
$\mathcal{K}$, the following formula holds
\begin{equation}
h(\{S^{t}\}) = \sqrt{-\mathcal{K}}
\end{equation}
and one usually puts the Gaussian curvature $\mathcal{K}=-1$.}
\begin{equation}
h(\{S^{t}\} \mbox{ on } PGL(2,\mathbb{Z})\backslash{\mathbb{H}}) =
1
\end{equation}
thus the billiard representation on the hyperbolic plane and the
BKL approach based on the Gauss map are not equivalent, as they
are not isomorphic as dynamical systems. In fact, the first
describes the behavior of a generic inhomogeneous singularity, the
second the fate of a Bianchi IX homogeneous universe. Yet, as we
mentioned above, a famous result due to Artin relates the
ergodicity of the geodesic flow on
PGL$(2,\mathbb{Z})\backslash{\mathbb{H}}$ to the ergodicity of the
Gauss map. Artin's theorem supports the conjecture that the
behavior of a generic space-like singularity is somehow well
described by a Bianchi IX homogeneous cosmological model,
conjecture for which nowadays there is a lot of numerical evidence
\cite{Berger}. At this point, we should add that some theoretical
progress has been made in understanding the general behavior of
the cosmological singularity thanks to C. Uggla et al (see
\cite{Uggla} for a review), who have shown that \emph{asymptotic
silence} holds, i.e. particle horizons along all timelines shrink
to zero for generic solutions. With Uggla's words at the last 11th
Marcel Grossmann Meeting, ``Everybody dies alone''.

We can make an interesting consideration when we pass from the
3-dimensional Minkowskian billiard to the 2-dimensional Euclidean
one. The light rays which move in the auxiliary Minkowskian space
have no mass of course. When we study the wave equation on the
reduced domain (i.e. the Schr\"{o}dinger equation for
PGL$(2,\mathbb{Z})$), there is a subtlety. Indeed, this equation
allows for a mass different from zero because it corresponds to
the quantization of a classically chaotic particle on the
hyperbolic plane (we always put $m=1$, $\hbar=1$). But this
particle ``arises'' from light-like rays in 3 dimensions. Thus, we
have an example of a mechanism to generate mass on a Euclidean
manifold from a null motion on a Minkowskian manifold.

\section{The Wave Function of the Universe}

Let us now come the quantization of the cosmological billiard. In
the language of \cite{DHN}, we can decompose the motion in the
3-dimensional billiard in an angular part inside a fundamental
domain for PGL$(2,\mathbb{Z})$ plus a radial part. The latter
gives a temporal contribution to the full wave function, because
the vertical direction in the Weyl chamber is a temporal one.
Thus, with the term wave function in this section we mean the
space part of the full wave function\footnote{To be more precise,
one should also say that the full wave function of the system is,
heuristically, the product of the wave functions in each space
point on the hyper-surface. But since the classical behavior is
local, we need only to study the quantum behavior in a generic
point, where we approximate the physical evolution by the billiard
representation.} which is solution of the Schr\"{o}dinger equation
on the hyperbolic plane. How to arrive at this equation after some
changes of variables is already described in many works (see
\cite{Montani} and references therein), we briefly remind the main
steps below. But let us stress that the wave
equation on the hyperbolic plane derives from the Wheeler-DeWitt
equation for the 3-dimensional problem. We believe that the
time-dependence of the solution to the full wave equation does not
change the physical spectrum, which is thus contained in the
2-dimensional billiard. In \cite{Montani} (following previous
works by Misner), the authors transform the domain for Bianchi IX
using an approximation which makes the domain integrable (they
modify the lower arc of the billiard with a horizontal segment,
but note that the latter is not a geodesic segment on the
hyperbolic plane) and with the same area. The same approach was
used for Bianchi IX in \cite{Graham}, where the authors identify
the fundamental domain on the hyperbolic plane
specific\footnote{They restrict to Bianchi IX without rotation of
axes. If one considers also the rotation of axes, the fundamental
domain is the one for $\Gamma_{0}(2)$\cite{Marcolli}, which is a
subgroup of PSL$(2,\mathbb{Z})$. It is reasonable for the
generic/inhomogeneous case to expect a generic discrete group, and
in fact PSL$(2,\mathbb{Z})$ occurs.} for this situation and try to
resolve the corresponding wave equation.
\begin{figure}[htbp]
\centering
  \includegraphics[scale=0.35]{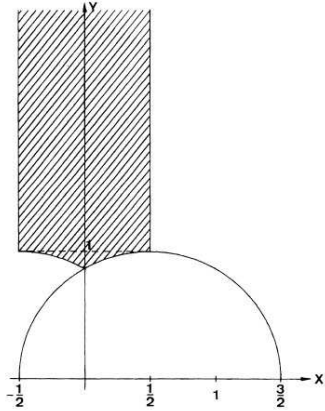}\\
  \caption{The domain on the hyperbolic plane corresponding to the asymptotic dynamics for a Bianchi IX universe
  (from \cite{Graham}).}
\end{figure}

Here we keep the \emph{exact} domain. The novelty with respect to
previous works is our widespread use of the arithmetic properties
of the billiard, which, although mentioned in some papers by DHN,
have not been taken in considerations before, especially at the
quantum level where they play a decisive role.

To be more specific, using the notation of \cite{Graham}, the 3-dimensional quantum equation
reads as follows:
\begin{equation}
-\frac{1}{\rho\,(\ln\rho)^{2}}\,\frac{\partial}{\partial\rho}[\rho\,(\ln\rho)^{2}]\frac{\partial\Psi}{\partial\rho}
+\frac{1}{\rho^{2}\,(\ln\rho)^{2}}[\Delta_{LB}-U(\rho,\zeta,\phi)]\Psi=0
\end{equation}
where $\Delta_{LB}=\frac{1}{\sinh\zeta}\,\frac{\partial}{\partial\zeta}(\sinh\zeta\,\frac{\partial}{\partial\zeta})
+ \frac{1}{\sinh^{2}\zeta}\frac{\partial^{2}}{\partial\phi^{2}}$, $U$ is a potential term
\begin{equation}
U(\rho,\zeta,\phi) = (\ln\rho)^{2}[\rho^{(4\cosh\zeta)}(V-1)]
\end{equation}
$\rho$ is the radial (time) variable, $\zeta$ and $\phi$ are the angular (space) variables.
This equation is exact, in the sense that it is derived from the
Bianchi IX metric without any other assumptions\footnote{In particular,
in the original Misner variables the Bianchi IX metric (\ref{bianchimetric}) can be written as
\begin{equation}
ds^{2}=-dt^{2} + \widetilde{a}^{2}(t)(e^{2\beta})_{ij}\omega^{i}\omega^{j}
\end{equation}
where $\beta=$diag$(\beta_{+}+\sqrt{3}\beta_{-},\beta_{+}-\sqrt{3}\beta_{-},-2\beta_{+})$ is 3 by 3 traceless matrix
and the $\omega_{i}$'s are a basis of one-forms for SO$(3)$. In terms of these variables,
the term $V$ in the potential $U$ is $V=\frac{1}{3}\,$Tr$(1-2\,e^{-2\beta}+e^{4\beta})$.
Finally, the variables $\beta_{\pm}, \widetilde{a}$ are transformed to hyperbolic coordinates
$\rho, \zeta, \phi$ \cite{Graham}.
}.

If we know take the limit $\rho \rightarrow 0^{+}$ (which corresponds to the case
$\Omega \rightarrow +\infty$ in the BKL variables), then the potential term
vanishes inside and is plus infinite
outside the triangular domain pictured in the figure. This makes it possible
to factorize the solution as follows
\begin{equation}
\Psi =\Phi(\rho) \psi(\zeta,\phi)
\end{equation}
and we consider only the equation
\begin{equation}
-\Delta_{LB}\psi_{i}=E_{i}\psi_{i}
\end{equation}
which after some redefinitions of variables corresponds to the spectral problem for the hyperbolic Laplacian
(see \cite{Montani} and \cite{Graham}, in this latter the approximate temporal evolution
for the 3-dimensional wave function is computed too).

To go beyond, as we have seen, the general inhomogeneous case is modelled on a billiard
problem inside the fundamental Weyl chamber of the hyperbolic
algebra HA$_{1}^{(1)}$ whose projection on the hyperbolic plane is
PGL$(2,\mathbb{Z})$ and not PSL$(2,\mathbb{Z})$, thus we have to
consider the spectral problem with Neumann boundary conditions
\begin{eqnarray}
  -\Delta \psi &=& E \psi \nonumber \\
  \psi & \in & L^{2}(\mathcal{F}_{3},\mu) \nonumber \\
  \partial_{n} \psi |_{\partial \mathcal{F}_{3}} & = & 0
\end{eqnarray}
where $\mathcal{F}_{3}$ is the halved standard modular domain,
$\Delta$ is the hyperbolic Laplacian and $\mu$ is the usual
measure on the hyperbolic plane (the notation of this section is
explained in Appendix \ref{maass} with more details). Thus
solutions to the Neumann problem for PGL$(2,\mathbb{Z})$ are given
by \emph{even} Maass cusp forms for PSL$(2,\mathbb{Z})$. In fact,
the spectrum of $\Delta$ on PSL$(2,\mathbb{Z})$ is purely discrete
on the space of odd functions, but on the even space there is a
continuous spectrum given by the interval
$\left[\frac{1}{4},\infty\right)$ and, what is less obvious, a
discrete spectrum. Thus the space of Maass waveforms (which for
PSL$(2,\mathbb{Z})$ are also cusp forms, i.e. they are zero at the
cusps) splits into odd/even forms under the symmetry $R_{1}$
\begin{equation}
\psi(-\overline{z}) = \pm \psi(z)
\end{equation}
Since the translation $T(z)=z+1$ belongs to the modular group, an
even/odd Maass form can be written as
$\psi(x)=\sum_{n}W_{n}(y)e^{2\pi i n x}$. Considering the
eigenfunction condition $\Delta \psi +
\left(\frac{1}{4}+t^{2}\right)\psi=0$ and the square-integrability
condition, one arrives at a more explicit form for the Maass cusp
forms for PSL$(2,\mathbb{Z})$, that is
\begin{equation}
\psi(z) = \sum_{n\neq 0}a_{\psi}(n)y^{1/2}K_{it}(2\pi  |n|
y)e^{2\pi i n x}
\end{equation}
where $K_{it}(2\pi  |n| y)$ are modified Bessel functions such
that $K_{it}(y) << e^{-2\pi y}$ for $y \rightarrow \infty$. The
coefficients $a_{\psi}(n)$ are the Fourier coefficients of the
$\psi(z)$. For even forms, $a_{\psi}(-n)=a_{\psi}(n)$, while for
odd ones $a_{\psi}(-n)=-a_{\psi}(n)$. Finally, since for
PSL$(2,\mathbb{Z})$ (and for all so-called Hecke triangle groups)
there is an obvious symmetry with respect to the imaginary axis,
one can write
\begin{equation}
\psi(z) =\sum_{n = 1}^{+\infty}a_{\psi}(n)y^{1/2}K_{it}(2\pi n
y)\left\{\begin{array}{ll}
  \cos\,(2\pi i n x) &  \\
  \sin\,(2\pi i n x) &  \\
\end{array}\right.
\end{equation}
depending on whether $\psi$ is even or odd. The precise statement,
then, is that \emph{the wave function of the universe is an even
Maass cusp form for the modular group} PSL$(2,\mathbb{Z})$. Apart
from the trivial eigenvalue $E_{0}=0$ and the corresponding
constant eigenfunction $\psi_{0}$, no other
eigenvalues/eigenfucntions are known analytically. Some eigenvalues, calculated via numerical investigations,
are given at the end of Section \ref{scarring_sec}.

Actually, more is true. In fact, since PSL$(2,\mathbb{Z})$ is
arithmetic and the cuspidal spectrum is very likely simple, one
can diagonalize the hyperbolic Laplacian and the Hecke operators
simultaneously. Thus it turns out that \emph{the wave function is
a Maass-Hecke eigenform}. This is an even more interesting
statement. In fact, Hecke eigenvalues are multiplicative, i.e.
$\lambda(mn) = \lambda(m)\lambda(n)$, and this should put some
conditions on the physical interpretation of them.

In recent years, in string theory, a lot of work has been done on
counting the entropy of black holes (see for example
\cite{Pioline} for a review). It turns out that in some cases this
entropy is counted by the Fourier coefficients of certain
automorphic functions. Thus, it is remarkable that the
quantization of gravity in the BKL limit leads to a wave function
which has automorphic properties. In the present case, the
question would be the following: \emph{is the gravitational
entropy computable from the Fourier coefficients of Maass-Hecke
eigenforms}?

Another remark is the following. In the semi-classical limit, the
expected statistics for the energy levels spacing distributions
for $X = PSL(2,\mathbb{Z})\backslash\mathbb{H}$ is the Poisson
distribution. Indeed, from general considerations regarding
chaotic systems, one would expect the system follow predictions
from random matrix theory (GOE/GUE ensemble). These ensembles
present level repulsion at short distance and rigidity at long
distance. But, again, due to the arithmetic property, we have an
anomalous statistics, the Poissonian one, which exhibits the
clustering property. The knowledge of the expected statistics
could, in principle, allow to compare with the observations.

The inclusion of matter changes the shape of the billiard and also
increases the number of dimensions of the billiard. The most
interesting case is perhaps supergravity theory in 11 dimensions,
which is believed to be a kind of ultimate theory unifying all
fundamental interactions. The supergravity billiard is the
fundamental Weyl chamber for E$_{10}$, thus in this case the
statement is that the wave function is a Maass waveform with
respect to the (discrete) Weyl group of E$_{10}$ (which is not
known). In this case, we can not speak safely of Maass cusp forms,
and we must use only the general term Maass waveform, because we
do not know if the residual spectrum is empty as in the cases of
each $\Gamma(N)$ (the congruence subgroups of
PSL$(2,\mathbb{Z})$). Besides, the discrete spectrum could also be
degenerate. The existence of Maass forms with respect to
W$($E$_{10})$ should be guaranteed by the fact W$($E$_{10})$ is
arithmetic.

Usually in quantum mechanics, one has a discrete and a continuous
spectrum for a self-adjoint Hamiltonian and these are separated.
Bound states are the proper eigenfunctions of the discrete
spectrum, whereas the continuous part is interpreted as a free
motion. Our model of quantum cosmology is different from this
usual situation. In fact, from the Selberg theory for the
automorphic Laplacian inside a fundamental domain of some
$\Gamma(N)$, we know that the discrete spectrum is \emph{not}
separated from the continuous one $[\frac{1}{4},\infty)$ (whose
improper eigenfunctions are given by the Eisenstein series), but
it is \emph{embedded} in the continuous part\footnote{There are
cases in scattering problems where points of the discrete spectrum
lie in the continuous spectrum, but our situation is
distinguished, because the \emph{whole} discrete spectrum is
embedded in the continuous one.}. We believe this says something
about the nature of quantum gravity, i.e. this suggests that
\emph{quantum gravity/cosmology is a non-trivial mixing of
discrete and continuous concepts}.

It would be interesting to understand if other approaches to the
problem of quantum gravity like the formalism of loop quantum
gravity or string theory say something regarding this point.

\section{Scarring and Asymptotic Number of Quantum States} \label{scarring_sec}

In \cite{BarrowLevin}, Barrow and Levin have analyzed the case of
a finite universe emerging from a compact octagon on the
hyperbolic plane. The octagon is a compact domain (but the
associated discrete group is not arithmetic), thus the classical
motion inside is chaotic. They found a fractal structure, which,
in the classical to quantum transition, can persist in forms of
scars, ridges of enhanced amplitude in the semi-classical wave
function. They conclude that if the universe is finite and
negatively curved, the cobweb of luminous matter might be a
residue of these primordial quantum scars.

Of course we can never know if our universe is finite or not. Yet,
in the case of a generic singularity, we have seen that the
interesting domain is the one for PGL$(2,\mathbb{Z})$, which is an
arithmetic group. Thus, \emph{there is no scarring} (see Appendix
\ref{maass}). The conclusions of Barrow and Levin do not apply in
this case and it remains to understand the physical (cosmological)
meaning for the absence of these scarred states. Results in this
direction by loop quantum gravity/cosmology or string theory would
be interesting, in order to confirm or discard the roles of
scarred stated in quantum cosmology.

Note one more thing. Our analysis is limited to the case of pure
gravity in 4 dimensions, thus the conclusion about the absence of
scarred states in quantum cosmology is valid only in this context.
Thus one may think that the inclusion of matter or extra
dimensions could change the situation. If we consider the case of
supergravity theory in 11 dimensions (or a not well defined
quantum version of it, like M-theory or whatever), then, as we
mentioned, the cosmological billiard is the Weyl chamber of
E$_{10}$ \cite{DHN} (the role of the fermions is still matter of
debate) whose Weyl group is still arithmetic. Thus for the
hyperbolic manifold $W($E$_{10})\backslash\mathbb{H}^{9}$ the
quantum unique ergodicity theorem should be true and, again, there
is no scarring effect. 11-dimensional supergravity is a candidate
theory to describe the universe and all of its interactions. The
message is that, with the knowledge we have today, it does seem
that \emph{in the early universe scarred states are absent}. And,
in our opinion, quantum gravity has to cope with quantum chaos.
The absence of scarred states is deduced from the arithmetic
quantum unique ergodicity theorem, which says that the only
invariant measure at the quantum level is the one given by the
Lebesgue measure on the hyperbolic manifold (which means
uniqueness of the invariant measure for this model of quantum
cosmology). In other words, all the features of classical chaos
disappear at the quantum level, giving a system which is closer to
a classically integrable system.

Finally, regarding the asymptotic number of quantum states, we can
count them using the Selberg result for PSL$(2,\mathbb{Z})$
\begin{equation}
N_{PSL(2,\mathbb{Z})}(R) = \sum_{0 < E_{j} \leq R}1\, \sim
\frac{\mu(X)}{4\pi}\,R
\end{equation}
for $R \rightarrow \infty$. This gives the asymptotic number of
eigenvalues/eigenfunctions, thus \emph{it counts the asymptotic
number of discrete states in quantum cosmology}. Actually,
we have to consider only the eigenvalues whose eigenfunctions are
even, so only a part (on average, half) of this asymptotics is physical. Again, a
confirmation of this from other approaches to the problem to
quantum cosmology would be extremely interesting. For E$_{10}$,
there is no such a result.

The first eigenvalues of $\Delta$ on X are $E_{1} =
91.12, E_{2} = 148.43, E_{3} = 190.13, E_{4}=206.16$. They are
respectively even, odd, odd, even and odd, with respect to the
symmetry about $x = 0$. See \cite{Hejhal} and \cite{Sarnak} for
the description of the numerical methods used to determine
eigenvalues and an approximate form for the eigenfunctions. It is
interesting also to compare the energy levels computed in
\cite{Montani} using the approximate domain with the ones computed
from the ``exact'' domain (but always via numerical methods).

\section{Conclusions, Some Speculations and Points to be developed in the future}

Assuming the billiard representation introduced by DHN in the
study of the cosmological singularity in the BKL limit, we have
commented on the question of integrability/chaoticity of the null
billiard flow inside the Weyl chamber of HA$_{1}^{(1)}$: one must
be careful about statements which derive the chaoticity of
cosmological billiard from the finite-volume argument (typical of
the Euclidean case), as we are in a pseudo-Riemannian setting. In
fact, from the Anosov property of the flow on the basis of the
billiard (the fundamental domain for PGL$(2,\mathbb{Z})$), it does
\emph{not} follow the chaoticity for the dynamics in the full
3-dimensional Weyl chamber, since from the ergodicity of a
dynamical system in a proper subset of the phase space one can not
infer the ergodicity of the system in the full phase space. We do
hope to study billiards and flows in pseudo-Riemannian manifolds
in the near future, since these are the natural evolution of
classical billiards towards considering the problem of ergodicity
in special/general relativity. The next step, in fact, should be
the study of relativistic billiards confined in some polyhedron in
a Minkowski spacetime.

We have also carried on the quantum analysis for the cosmological
billiard using the arithmetic properties of the modular group. The
result is that, for pure gravity in 4 dimensions, the wave
function of the universe (or better, the angular part projected on
the hyperbolic plane, i.e. the space part of the full wave
function) is an automorphic form for PGL$(2,\mathbb{Z})$,
precisely an even Maass cusp form for PSL$(2,\mathbb{Z})$. Indeed,
since PSL$(2,\mathbb{Z})$ is arithmetic, \emph{the wave function
is a Maass-Hecke eigenform}, being also eigenfunction of the Hecke
operators (which commute with the Hamiltonian). In our opinion,
the most important prediction/hint of the model is that \emph{the
discrete spectrum is fully embedded in the continuous spectrum},
which is a very distinguished situation. In fact, for every point
in the discrete spectrum and every normalizable even Maass cusp
form, there exists a corresponding not-normalizable Eisenstein
series with the same value of the generalized eigenvalue. This is
an odd situation, thus we would like to understand if this
phenomenon hides some deep truth which is an intrinsic property of
quantum gravity/cosmology or if it is due to the billiard model
which is too simplified.

The arithmetic nature of PGL$(2,\mathbb{Z})$ allows to state
that in the early universe \emph{scarred states are absent},
thanks to the quantum unique ergodicity theorem by Lindenstrauss.
The model allows also to count the asymptotic number of quantum
states, whose number is half of the full Selberg asymptotics for PSL$(2,\mathbb{Z})$,
and to identify a distribution for the level spacing
statistics. This distribution is the Poissonian one and is
characterized by the clustering property. One should understand
the meaning of this as opposite to the level repulsion predicted
from random matrix theory for non-arithmetic chaotic systems.

These conclusions, true for pure gravity in 4 dimensions if we
believe the billiard representation, are very likely valid also in
the case of 11-dimensional supergravity close to the singularity,
whose billiard is modelled on the E$_{10}$ hyperbolic Kac-Moody
algebra. Very briefly, one can say that the model predicts
specific automorphic properties for the wave function, it gives
the asymptotic number of quantum states and the statistics for the
level spacing. Arithmetic systems at the quantum level seem to be
closer to integrable systems than to generic chaotic systems,
thus, for this model, any manifestation of classical chaos
disappears at the quantum level.

In \cite{Forte}, we have pointed out that the Selberg trace
formula for PSL$(2,\mathbb{Z})$ gives a semiclassical quantization
rule for pure gravity in 4 dimensions, as typically occurs for the
Gutzwiller trace formula in quantum chaos. But the difference is
that the Gutzwiller trace formula is divergent, whereas the
Selberg trave formula is convergent. Thus, in the BKL limit to the
singularity and in the billiard representation, a semiclassical
quantization of gravity is well defined. Indeed, the Selberg trace
formula is a kind of path-integral. This point of view is
emphasized in the book by C. Grosche \cite{Grosche}. But remember
that indeed we have an infinite number of semi-classical
quantization rules, since the class of test functions entering the
trace formula is very large \cite{Sarnak}. This is the best one
can do in a rigorous/convergent way for the quantum systems whose
semi-classical limit is a Hamiltonian flow for which the Selberg
trace formula is valid. Nevertheless, the trace formula contains a
lot of information, in fact Selberg derived his asymptotics using
a particular Gaussian test function.

It may be that this quantization helps in finding a correct theory
of quantum gravity, as it happened in the early days of quantum
mechanics after the introduction of Borh-Sommerfeld quantization
rules. The Selberg trace formula, in a sense, is an identity
between quantum mechanics (because it contains the spectrum of the
quantum Hamiltonian) and classical mechanics (because it contains
a sum over all classical periodic orbits). In \cite{Forte}, we
have made an attempt to interpret the imaginary roots of the
hyperbolic algebra in terms of the periodic orbits which enter the
Selberg trace formula. These periodic orbits live inside
PSL$(2,\mathbb{Z})$, which is the projected billiard. The idea is
to check if these periodic orbits on the hyperbolic plane can be
lifted up to periodic orbits (which escape the singularity) in the
3-dimensional billiard such that they correspond to solutions
(also approximate) of Einstein equations in the asymptotic regime
as it happens for the Kasner solutions. In fact, to the simple
roots of the algebra and their Weyl-images (i.e. the real roots)
one can associate Kasner solutions, because every reflection is a
Weyl reflection from a Kasner epoch to another one. We would like
to understand if something similar is possible for the imaginary
roots, in particular relating the Selberg trace formula to the
structure of the hyperbolic algebra. A future paper of more
mathematical taste is in progress.

The billiard representation is very interesting from many points
of view. It is evident that in this formalism one has another
example of a relation between general relativity and geometrical
optics. One can wonder if this reformulation can be applied to
other models too. We would like to understand if it is possible to
reformulate problems of general relativity as billiard-like
problems (although billiard problems are generally very difficult
to solve). This would help in studying the integrability or
ergodicity of general relativity in particular situations.

For example, the variables used to describe Choptuik's solution to
the collapse of a spherically symmetric scalar field show a
profile with spikes which resemble (billiard) bounces, see for example
figure 3 in \cite{Gundlach}\footnote{The author thanks prof. T
Damour for explaining Choptuik's discover to him.}. Moreover, this
system exhibits discrete self-similarity, a property which is very
likely shared by the BKL limit too (let us remember that the
collapse studied numerically by Choptuik implies the formation of
a black-hole, that is a singularity, thus possible analogies with
the cosmological singularity are reasonable). However, let us remember that
Choptuik's solution is physical, while the BKL behavior, as far as we know,
is theoretical.

We have limited our analysis to the case of pure gravity in 3+1
dimensions, where, in the limit to the singularity, spatial points
decouple and the dynamics involves only a single (time) variable.
The question now is: what could be the analog for a quantum theory
when all 4 (time \emph{and} space) variables must be considered?
Is it possible to describe gravity with the tools of higher-dimensional
modular groups in particular regimes where one does not have spatial
decoupling? Of course, this is a statement motivated more by beautiful mathematics
than physics, thus it can be wrong. However, let us remember that
higher-dimensional groups share with PSL$(2,\mathbf{Z})$ a lot of
nice properties, like existence of waveforms, trace formulae (the
so-called Arthur trace formula) etc. In other words, the construction
of a quantum theory would be viable under the the aegis of what we would call
an ``automorphic principle''\footnote{Note that this proposal is different from
the one contained for example in \cite{DHN} (and references
therein). DHN's approach is based on the assumption (for which
there is indeed evidence) that supergravity theory in 11
dimensions (or a quantum extension of it) exhibits a hidden
E$_{10}$ symmetry at the level of the Lagrangian, i.e. there
should exist a formulation of this theory invariant under the
infinite-dimensional Lie group E$_{10}\backslash$K(E$_{10})$,
where K(E$_{10})$ is the formal maximal compact subgroup of the
E$_{10}$ Lie group. In this setting, the discrete group one looks
for to build the quantum theory is a discretized version of
E$_{10}$, E$_{10}(\mathbb{Z})$. Our proposal is different: we
suggest to look at discrete groups like
GL$(n,\mathbb{Z})$ and subgroups, \emph{not} to some HA$_{1}^{(1)}(\mathbb{Z})$.
Building automorphic forms for GL$(n,\mathbb{Z})$ groups (and
subgroups) is a well-established technique, while automorphic
forms for E$_{10}(\mathbb{Z})$ or HA$_{1}^{(1)}(\mathbb{Z})$, at
the moment, are out of reach. Anyway, all these considerations
can have nothing to do with true physics.}. An interesting mathematical
mechanism which could do the job is the lifting of automorphic forms
to higher dimensions, but again the connection with the physics is all to prove.
Everything is mathematically well defined and beautiful, and leads,
unavoidably, to the classical Langlands program. Very recently, E.
Witten et al have given a physical interpretation of the geometric
Langlands program in terms of gauge theory. This construction uses
a lot Hecke eigensheaves, whose classical counterpart is Hecke
operators and Hecke eigenfunctions. It is remarkable, in my
opinion, that these objects appear in general relativity in a
different context and with a different language. It could be
another subtle indication of the so-called gauge-gravity
correspondence, realized via primitive (arithmetic) objects.

\section{Acknowledgements}

The author wishes to thank profs. T. Damour, F. Englert, L.
Houart, M. Henneaux for fruitful discussions at early stages of
this work, prof. G. Esposito for reading a copy of the manuscript and prof. A. Sciarrino for
suggestions and collaboration in the years. Finally, the author wishes to thank the referees
for their useful comments and for pointing out imprecisions and/or improvements.

\appendix

\section{The Hyperbolic Kac-Moody Algebra HA$_{1}^{(1)}$} \label{kacmoody}

In this section we do not review the theory of Kac-Moody algebras
(see \cite{Kac} for details), but just fix the notation.

The algebra HA$_{1}^{(1)}$ is an infinite-dimensional Lie algebra,
being a hyperbolic Kac-Moody algebra. This means that it is
identified by the  following (generalized) Cartan matrix
\begin{equation}
A=a_{ij}=\left(%
\begin{array}{ccc}
  2 & -2 & 0 \\
  -2 & 2 & -1 \\
  0 & -1 & 2 \\
\end{array}%
\right)
\end{equation}
or equivalently by its Dynkin diagram
\begin{center}
 \begin{picture}(40,20) \thicklines
        \put(0,0){\circle{14}}
        \put(42,0){\circle{14}}
        \put(-42,0){\circle{14}}
        \put(0,15){\makebox(0,0){$\alpha_{2}$}}
        \put(42,15){\makebox(0,0){$\alpha_{1}$}}
        \put(-42,15){\makebox(0,0){$\alpha_{3}$}}
        \put(5,-3){\line(1,0){31}}
        \put(5,3){\line(1,0){31}}
        \put(-7,0){\line(-1,0){28}}
        \put(35,0){\line(-1,1){10}}\put(35,0){\line(-1,-1){10}}
        \put(7,0){\line(1,-1){10}}\put(7,0){\line(1,1){10}}
 \end{picture}
\end{center}
The precise definition of the algebra is standard and goes as
follows. Let $\mathfrak{h}$ be a complex vector space whose
dimension is 3, and $\mathfrak{h}^{\ast}$ its dual. Then there
exist linearly independent indexed sets
\begin{equation}
\Pi := \{\alpha_{i}\} \subset \mathfrak{h}^{\ast}\,,\quad
\mbox{and } \Pi^{\vee} := \{h_{i}\} \subset \mathfrak{h}\,,
\end{equation}
such that $\alpha_{j}\,(h_{i}) = a_{ij}$ ($i,j=1,2,3$). The
$\alpha_{i}$ ($h_{i}$) are called \emph{simple roots} (\emph{dual
simple roots}). The sets $\Pi$ and $\Pi^{\vee}$ are uniquely
determined by $A$ up to isomorphism.

Then the \emph{Kac-Moody algebra} HA$_{1}^{(1)}$ is the complex
Lie algebra generated by $\mathfrak{h} \cup \{e_{i},f_{i}\}$ with
the following defining relations:
\begin{equation}
[e_{i},f_{j}]=\delta_{ij}\,h_{i}\,,\quad [h,h^{\prime}]=0 \mbox{
for } h,h^{\prime} \in \mathfrak{h}
\end{equation}
$$
[h_{i},e_{j}] = a_{ij}\,e_{i}\,,\quad [h_{i},f_{j}] = -
a_{ij}\,f_{i}
$$
$$
(\mbox{ad }e_{i})^{1-\,a_{ij}}\,e_{j} = 0\,,\quad (\mbox{ad
}f_{i})^{1-\,a_{ij}}\,f_{j} = 0\,\mbox{ for } i \neq j\,.
$$
$\mathfrak{h}$  is the maximal abelian subalgebra of
$\mathfrak{g}(A)$ (\emph{Cartan subalgebra}). The order of the
Cartan matrix is also called the \emph{rank} of the algebra, which
in this case coincides with the rank of the matrix $A$. Finally,
the algebra is simple.

The term hyperbolic refers to the fact that its Cartan matrix has
determinant $-1$ and Lorentzian signature $(-++)$. In a sense, it
is the simplest hyperbolic Kac-Moody algebra of rank 3, being the
canonical hyperbolic extension of A$_{1}$ ($\mathfrak{su}(2)$)
trough the affine node $\alpha_{2}$ and the hyperbolic node
$\alpha_{3}$. The scalar products between the simple roots
$(\alpha_{i},\alpha_{j})=a_{ij}$ are
\begin{equation}
(\alpha_{1},\alpha_{2})=-2\,,\quad(\alpha_{1},\alpha_{3})=0\,,\quad(\alpha_{2},\alpha_{3})=-1
\end{equation}

$A$ is a flat metric on a certain pseudo-Riemannian vector space
$\mathfrak{h}_{\mathbb{R}}^{\ast}$ (see below). Given the simple
roots, one has the notion of real roots, which are $W-$equivalent
to the simple roots, i.e. they are the image through the Weyl
group of the simple roots, and imaginary roots. It turns out that
the real roots $\alpha$ have positive squared norm, while for
imaginary roots $\alpha^{2} \leq 0$. Indeed, for hyperbolic
(symmetrizable) Kac-Moody algebras, the imaginary roots are
precisely all the vectors in the root lattice
$\bigoplus_{i}\mathbb{Z}\alpha_{i}$ which have zero or negative
squared norm (Moody theorem).

The Weyl group $W$ is defined by the fundamental reflections in
the simple roots
\begin{equation}
R_{i}(\beta) = \beta -
2\,\frac{(\alpha_{i},\beta)}{(\alpha_{i},\alpha_{i})}\,\alpha_{i}
\end{equation}
It is a Coxeter group, i.e. it is a discrete group generated by
the fundamental reflections with the following relations
\begin{equation}
R_{i}^{2}=1\,,\quad \quad  \left(R_{i}R_{j}\right)^{m_{ij}}=1
\end{equation}
where the Coxeter exponents $m_{ij}$ are positive integers or
$\infty$ (in this case one puts $x^{\infty}=1$). For Kac-Moody
algebras, one has $m_{ij}=2,3,4,6$ or $\infty$ according to
$a_{ij}\,a_{ji}=0,1,2,3$ or $\geq 4$. Thus, different Kac-Moody
algebras with the same Coxeter exponents has isomorphic Weyl
groups\footnote{For example, the two rank-3 algebras
\begin{equation}
\left(%
\begin{array}{ccc}
  2 & -1 & 0 \\
  -1 & 2 & -4 \\
  0 & -1 & 2 \\
\end{array}%
\right),\quad
\left(%
\begin{array}{ccc}
  2 & -1 & 0 \\
  -1 & 2 & -1 \\
  0 & -4 & 2 \\
\end{array}%
\right)
\end{equation}
have the same Weyl group as HA$_{1}^{(1)}$.}. For HA$_{1}^{(1)}$
\begin{equation}
R_{i}^{2}=\left(R_{1}R_{3}\right)^{2}=\left(R_{2}R_{3}\right)^{3}=1
\end{equation}
and the Weyl group can be easily identified with
PGL$(2,\mathbb{Z})$ putting $T=R_{2}R_{1}$ and $S=R_{1}R_{3}$,
where $T, S$ are the standard generators for PSL$(2,\mathbb{Z})$.
The even (positive) Weyl group $W^{+}$ is thus isomorphic to
PSL$(2,\mathbb{Z})$. The properties of PSL$(2,\mathbb{Z})$, the
modular group, are recalled in the next section.

Every Coxeter group defines a Coxeter polytope, which is the
polyhedron whose faces are point-wise fixed by the fundamental
reflections. When the Coxeter group is the Weyl group of a
Kac-Moody algebra, this polyhedron is called the Weyl
chamber\footnote{To be honest, the Weyl chamber is the domain in
$\mathfrak{h}$ defined by
\begin{equation}
\{h \in \mathfrak{h}_{\mathbb{R}}|\,\alpha_{i}(h)\geq 0\}
\end{equation}
where $\mathfrak{h}_{\mathbb{R}}=\{h \in
\mathfrak{h}|\,\alpha_{i}(h) \in \mathbb{R}\}$ is a $W-$stable
real subspace of $\mathfrak{h}$. In this notation, what we call
Weyl chamber should more correctly called the dual Weyl chamber,
since we are interested in the polyhedron in
$\mathfrak{h}_{\mathbb{R}}^{\ast}$.} of the algebra and is
contained in $\mathfrak{h}_{\mathbb{R}}^{\ast}$ . Thus the faces
of the Weyl chamber are orthogonal (with respect to the metric
given by the Cartan matrix) to the simple roots and are
hyperplanes in the pseudo-Riemannian space
$\mathfrak{h}_{\mathbb{R}}^{\ast}$.

\section{Maass Automorphic Forms for PSL$(2,\mathbb{Z})$}\label{maass}

Let us remind the main properties of the modular group
PSL$(2,\mathbb{Z})$. First we consider PGL$(2,\mathbb{Z})$,
sometimes called the extended modular group, whose fundamental
domain we call $\mathcal{F}_{3}$ (remember that one of the angles
is $\pi/3$). It is a discrete group acting on the hyperbolic plane
and is generated by the hyperbolic reflections in the three sides
\begin{equation}
R_{1}(z) = -\overline{z}\,,\quad R_{2}(z) = -\overline{z} +
1\,,\quad R_{3}(z)=\frac{1}{\overline{z}}
\end{equation}
PSL$(2,\mathbb{Z})$ is its most important subgroup, its domain
$\mathcal{F}$ is twice the fundamental domain of
PGL$(2,\mathbb{Z})$
\begin{equation}
\mathcal{F} = \left\{z \in \mathbb{H}:|z|>1, |x|\leq \frac{1}{2}
\right\}
\end{equation}
and its hyperbolic area is
\begin{equation}
\mu(\mathcal{F}) = \frac{\pi}{3}
\end{equation}
where $\mu$ is the usual hyperbolic measure on the hyperbolic
plane, $d\mu = dx\,dy/y^{2}$. The standard generators for
PSL$(2,\mathbb{Z})$ are
\begin{equation}
T = \left(%
\begin{array}{cc}
  1 & 1 \\
  0 & 1 \\
\end{array}%
\right)\,,\quad S = \left(%
\begin{array}{cc}
  0 & 1 \\
  -1 & 0 \\
\end{array}%
\right)
\end{equation}
considered as fractional linear transformations, i.e.
\begin{equation}
T(z) = z + 1\,,\quad S(z) = -\,\frac{1}{z}
\end{equation}
$S$ is such that $S^{2}=1$ ($S=S^{-1}$); moreover Tr $S=0 < 2$, Tr
$T=2$ thus $S$ is an elliptic element, $T$ is a parabolic element
(for details on hyperbolic geometry and Fuchsian groups see
\cite{Katok}).

\begin{figure}
\centering
  \includegraphics[width=6cm,keepaspectratio]{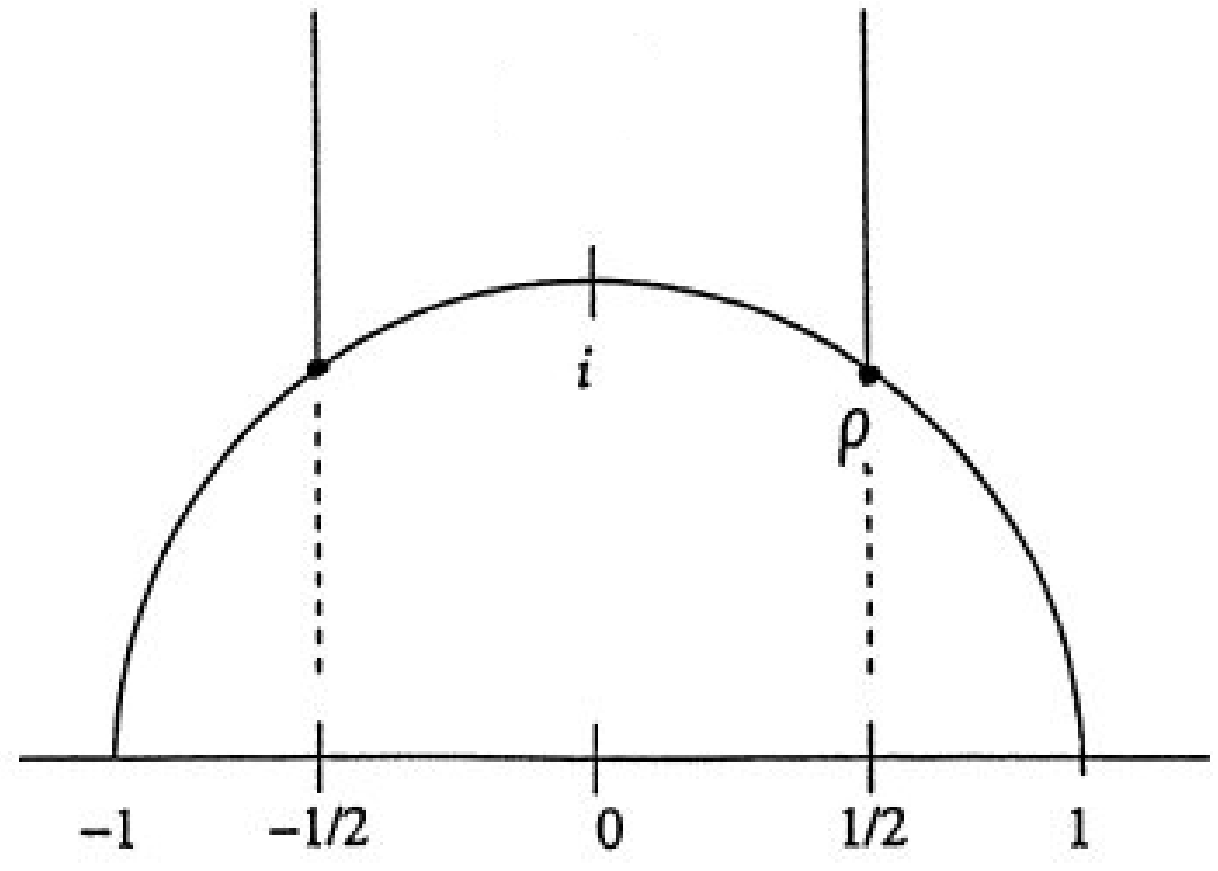}%
  \includegraphics[width=6cm,keepaspectratio]{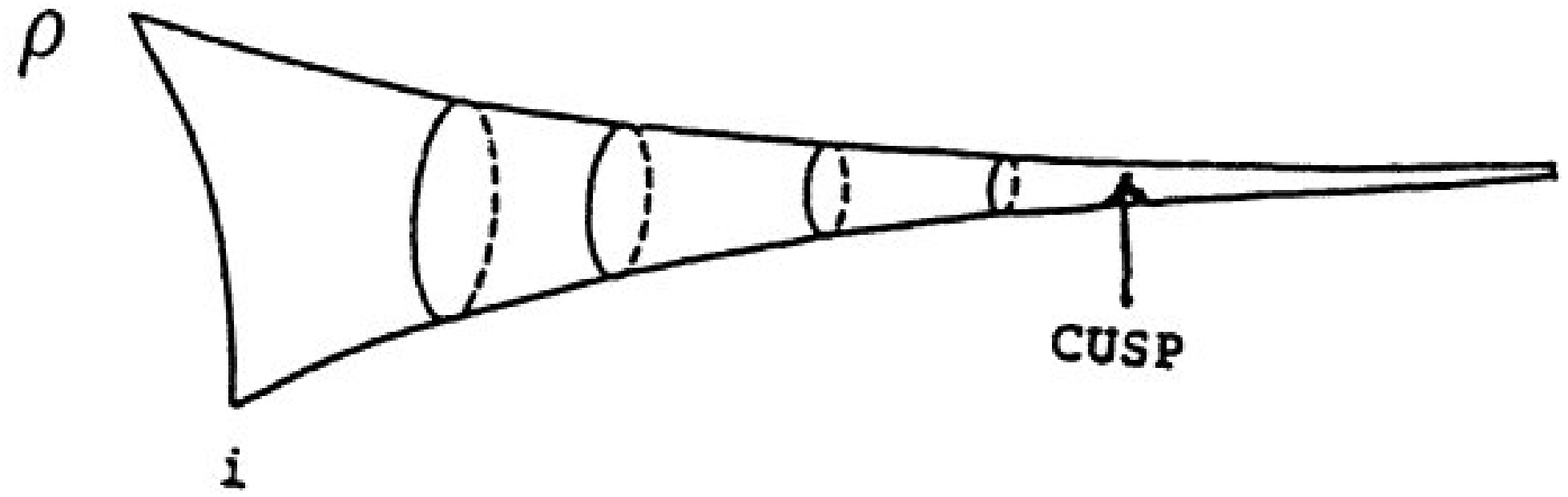}
  \caption{The modular domain (the domain for PGL$(2,\mathbb{Z})$ is the desymmetrized one) and the modular surface (from \cite{Sarnak}).}
\end{figure}

$T$ and $S$ can be expressed of course in terms of the reflections
$R_{i}$ belonging to PGL$(2,\mathbb{Z})$
\begin{equation}
T = R_{2}R_{1} \neq R_{1}R_{2} = T^{-1}\,,\quad S=
R_{1}R_{3}=R_{3}R_{1}
\end{equation}
Note that PGL$(2,\mathbb{Z})$ contains hyperbolic reflections,
that is negative isometries with determinant $-1$, while
PSL$(2,\mathbb{Z})$ contains only positive isometries. Thus we can
consider the hyperbolic surface $X:=$
PSL$(2,\mathbb{Z})\backslash\mathbb{H}$, called the \emph{modular
surface}, which is an oriented, finite area, non-compact
2-dimensional Riemannian manifold. It is also a Riemann surface
with the complex structure inherited by $\mathbb{H}$. It has a
cusp at infinity, which corresponds to the fixed point $i\infty$
of the parabolic transformation $T$. We are interested in the
spectral problem for $X$. See the excellent survey \cite{Sarnak}
for more details and references.

The so-called fundamental spectral problem of quantum chaos is the
following
\begin{equation}
\left \{
\begin{array}{ll}
 \Delta\psi + E\psi = 0\,, & E=\frac{1}{4} + t^{2}>0 \\
 \psi(\gamma z) = \psi(z)\quad \forall\; \gamma \in \mbox{ PSL}(2,\mathbb{Z}) &  \\
 \int_{X}|\psi(z)|^{2}d\mu(z) < +\infty &
\end{array}
\right.
\end{equation}
Here $\Delta=y^{2}(\partial^{2}_{x} + \partial^{2}_{y})$ is the
hyperbolic Laplacian. The numbers $0 = E_{0} < E_{1} \leq E_{2}
\leq \cdots$ for the which the spectral problem has solutions form
the discrete spectrum (eigenvalues) for $X$. The only known
eigenvalue is $E_{0}=0$, to which it corresponds the constant
eigenfunctions $\psi_{o}(z)$ (remember that $\mathcal{F}$ has
finite area). Solutions to the previous spectral problem are
called \emph{Maass automorphic forms} (or Maass waveforms), after
the mathematician H. Maass, who first studied them. The term
automorphic refers to the periodicity under PSL$(2,\mathbb{Z})$,
i.e. $\psi(\gamma z)= \psi(z)$. It is not evident that solutions
to this problem exist, since the modular surface is not compact.
Moreover, no explicit eigenvalues or eigenfunctions are known or
expected for $X$, although Maass himself produced an explicit
subsequence of eigenvalues for other discrete groups (e.g.
$\Gamma(4)$).

One would like to formulate for this case the analog of the Weyl
law for the Dirichlet problem for the Euclidean Laplacian on a
compact domain in $\mathbb{R}^{2}$ on the asymptotic number of
eigenvalues. This is possible thanks to Selberg theory, in
particular his famous trace formula. Selberg found that there
exists a continuous spectrum too, and the latter is the whole
interval $\left[\frac{1}{4}\,, \infty\right)$ with multiplicity
one. The corresponding not-normalizable eigenfunctions are given
by the Eisenstein series, which, for $X$, read as follows
\begin{equation}
E(z,s) = \sum_{g\,\in\, \Gamma_{\infty}\backslash
PSL(2,\mathbb{Z})} \frac{y^{s}}{|cz+d|^{s}}\quad \mbox{ for Re }s
> 1
\end{equation}
where $\Gamma_{\infty}$ is the stabilizer of $\infty$, which is
the only cusp, $z=x+iy$ and the transformation $g$ is represented
as usual by the matrix $\left(%
\begin{array}{cc}
  a & b \\
  c & d \\
\end{array}%
\right)$. The Eisenstein series extend meromorphically to the
whole $\mathbb{C}-$plane and are analytic on Re $s = \frac{1}{2}$.
The continuous spectrum is thus furnished by these generalized
eigenfunctions $E\left(z,\frac{1}{2}+it\right)$, $t\geq 0$,
\begin{equation}
\Delta E\left(z,\frac{1}{2}+it\right) + \left(\frac{1}{4} +
t^{2}\right)\,E\left(z,\frac{1}{2}+it\right) = 0
\end{equation}
and these are of course PSL$(2,\mathbb{Z})-$periodic
\begin{equation}
E(gz,s) = E(z,s) \mbox{ for any }g \in PSL(2,\mathbb{Z})
\end{equation}
that is they are generalized Maass automorphic forms. Let us call
$\overline{\psi}(s)$ the constant term in the Fourier expansion of
the Eisenstein series: this is meromorphic in $\mathbb{C}$ and for
Re $s \geq 1/2$ its poles are in $\left(\frac{1}{2},1 \right]$.
The residues at these poles are solutions to the spectral problem
and form the residual spectrum of $X$. If we now take the
orthogonal complement in $L^{2}(X,\mu)$ of the continuous and
residual spectrum, we obtain the cuspidal space $L^{2}_{cups}(X)$.
It is invariant under the hyperbolic Laplacian and the resolvent
$(\Delta-\lambda)^{-1}$ is compact when restricted $L^{2}_{cusp}$.
A Maass form which also lies in $L^{2}_{cusp}$ is called a
\emph{Maass cusp form}. These cusp forms are a kind of building
blocks for the theory of automorphic forms. Their existence is of
course tied to the size of $L^{2}_{cusp}$, since
$L^{2}_{cusp}\neq\{0\}$ is not obvious.

However, it turns out that for PSL$(2,\mathbb{Z})$,
$\overline{\psi}(s)$ has no poles in $\left(\frac{1}{2},1
\right)$, thus the residual spectrum is empty and any Maass form
is automatically a cusp form. Using his trace formula and the fact
that for PSL$(2,\mathbb{Z})$ $\overline{\psi}(s)$ can be expressed
in terms of a ratio involving the (completed) Riemann
zeta-function, Selberg was able to show that the contribution of
the continuous spectrum to the Weyl law is negligible, and that
\begin{equation}
N^{cusp}_{PSL(2,\mathbb{Z})}(R) : = \sum_{0 < E_{j} \leq R} 1 \sim
\frac{\mu(\mathcal{F})}{4\pi}\,R\quad (R \rightarrow \infty)
\end{equation}
that is solutions to the spectral problem exist and in abundance
(each eigenvalue is counted with its multiplicity). Let us stress
that no eigenvalues/eigenfunctions are explicitly known.

The existence of Maass cusp forms for PSL$(2,\mathbb{Z})$ is
deeply related to the its \emph{arithmetic}
nature\footnote{Phillips and Sarnak have more generally shown that
the arithmetic groups are the only ones which allow for solutions.
Many numerical experiments support the absence of eigenvalues and
Maass forms for non-arithmetic cases. For the precise definition
of an arithmetic Fuchsian group see \cite{Katok}.}. One can also
consider the spectral problem for PGL$(2,\mathbb{Z})$; this latter
contains negative isometries, thus one can not form a hyperbolic
manifold by taking the quotient, but one can consider the spectral
problem with Neumann boundary conditions
\begin{eqnarray}
  -\Delta \psi &=& E \psi \nonumber \\
  \psi & \in & L^{2}(\mathcal{F}_{3},\mu) \nonumber \\
  \partial_{n} \psi |_{\partial \mathcal{F}_{3}} & = & 0
\end{eqnarray}
Solutions to this problem are given by the even Maass cusp forms
for PSL$(2,\mathbb{Z})$.

In full generality, let $X(N)$ be the surfaces built out of the congruence subgroups
$\Gamma(N)$. The spectral problem applies exactly as above. Regarding the low-energy spectrum of $X(N)$,
let us call $E_{1}(N)$ the
closest eigenvalue to $E_{0}=0$. A deep conjecture (still open) of
Selberg states that
\begin{equation}
E_{1}(N) \geq \frac{1}{4}
\end{equation}
For $N=1$ ($X\equiv X(1)$), the case of the modular group, the numerical calculations
show that the first eigenvalue is $E_{1}=91.12\ldots$.
Since there is no residual spectrum\footnote{besides $E=0$, which thus should not be considered in the discrete spectrum.},
we can take for $E_{1}$ the
smallest eigenvalue of a Maass cusp form on $X$. One can
understand the number $1/4$ by recalling that a result due to
McKean shows that the spectrum of $\Delta$ on the universal
covering $L^{2}(\mathbb{H})$ is $\left[\frac{1}{4},\infty
\right)$. Moreover, P. Cartier conjectured that the cuspidal
spectrum of $\Delta$ on $X$ is \emph{simple} (the situation may indeed be very different for the other surfaces $X(N)$), a result confirmed
by many numerical experiments. For the high-energy behavior of the
spectrum, one would expect, using standard arguments from random
matrix theory, that it would fit the GOE-ensemble predictions (the
system is time-invariant), as the geodesic flow on $X$ is an
Anosov flow (thus chaotic). Yet, all the numerical experiments
show that the high-energy spectrum follows a Poissonian
distribution (which is instead typical of systems whose
semi-classical limit is integrable). The reason for that is,
again, the arithmetic nature of PSL$(2,\mathbb{Z})$, in particular
the existence of additional symmetries, the so-called \emph{Hecke
operators}, which for $n
> 0$ read as follows
\begin{equation}
T_{n}\psi(z) : = \frac{1}{\sqrt{n}}\,\sum_{ad=n,\;b \mod
n}\psi\left(\frac{az+b}{d} \right)
\end{equation}
the sum going over all positive integers $a, d$ with $ad=n$ and
$0\leq b<d$. These operators are a commutative
family\footnote{Mathematicians have no doubts that the anomalous
statistics for arithmetical dynamical systems is due to these
Hecke operators, which at the quantum level commute with the
Laplacian $\Delta$ and somehow mimic an integrable system. A
rigorous proof of that is lacking.} of self-adjoint operators on
$L^{2}(X)$ and commute with $\Delta$ and the reflection $R_{1}$. Thus they preserve the even/odd
eigenspaces of $\Delta$ and each eigenspace has a basis consisting
of simultaneous eigenfunctions of all Hecke operators. Such
eigenfunctions are called \emph{Maass-Hecke eigenforms}. Given
such an eigenfunction $\psi$, $T_{n}\psi = \lambda_{n}\psi$, its
Fourier coefficients are given by
\begin{equation}
a_{\psi}(n)= a_{\psi}(1)\lambda_{n}
\end{equation}
and we can normalize the first Fourier coefficient
$a_{\psi}(1)=1$; in this way the $n-$th Fourier coefficient is the
Hecke eigenvalue $\lambda_{n}$. Hecke eigenvalues enjoy a lot of
nice properties, the most important is that they are
multiplicative
\begin{eqnarray}
  \lambda_{mn} &=& \lambda_{m}\,\lambda_{n}\quad m,n\mbox{ co-prime}  \\
  \lambda_{p^{k}}\lambda_{p} &=& \lambda_{p^{k+1}} +
  \lambda_{p^{k-1}}\quad p\mbox{ prime}
\end{eqnarray}
Let us conclude this section by recalling a deep and recent
mathematical result about the quantum unique ergodicity for
arithmetic hyperbolic manifolds. Let us first observe that, from
the classical point of view, arithmetical dynamical systems are
chaotic as any other model on compact negatively curved manifolds.
But at the quantum level the arithmetic property is peculiar. If
one defines the probability measures $\mu_{j}$
\begin{equation}
d\mu_{j} = |\psi_{j}|^{2}d\mbox{vol}
\end{equation}
where $\psi_{j}$ is an eigenfunction of the Laplacian and $d$vol
is the Riemannian volume element, then the quantum unique
ergodicity conjecture by Rudnick and Sarnak states that for
compact manifolds of negative curvature, the measures $\mu_{j}$
converge to $d$vol (in the weak$^{\ast}$ topology). The conjecture
by Rudnick and Sarnak, if true, is remarkable, because it asserts
that at quantum level and in the semi-classical limit, there is no
manifestation of chaos. In particular, one would have quantum
unique ergodicity, that is only one possible quantum limit,
whereas classical unique ergodicity, i.e. uniqueness of the
invariant measure for the Hamiltonian flow, is never satisfied for
chaotic systems.

This conjecture is motivated by a theorem due to Shnirelman, which
says that for an ergodic system there exists a subsequence $j_{k}$
for which the measures $\mu_{j_{k}}$ converge to the standard
normalized Lebesgue measure. Shnirelman's theorem is an statement
of equidistribution, that is one often says that the
eigenfunctions equidistribute because the probability densities
$|\psi_{j_{k}}|$ tend to a constant independent on any point on
the manifold. Thus, for non exceptional $E_{n}$, $|\psi_{n}|^{2}$
can never localize to just a finite number of closed geodesics.
This theorem has been improved by Colin de Verdiere and Zelditch,
who showed that eigenfunctions still equidistribute on non-compact
manifolds like PSL$(2,\mathbb{Z})\backslash\mathbb{H}$. The
presence of an exceptional set is of course troubling. In
stadium-like domains, the so-called scarring effect has been
observed. For numerous $n$, the topography of $\psi_{n}$ is found
to contain clear ridges of mass of scars, situated roughly along
what would appear of closed geodesics. The location of these scats
changes with $n$. In practice, scars is what is left of periodic
orbits \cite{Hejhal}.

But for negatively curved manifolds, the QUE conjecture by Rudnick
and Sarnak denies the existence of these scars. The proof of this
conjecture is still lacking, yet progress has been made for
\emph{arithmetic hyperbolic surfaces} thanks to E. Lindenstrauss.
He has shown that for a compact arithmetic quotient the quantum
unique ergodicity conjecture is true and that, moreover, the same
statement holds for the modular surface (which is non-compact).
Thus, in the latter case, there is no scarring, i.e. these is no
manifestation of classical chaos at the quantum level.

\end{document}